\documentclass[reprint,groupedaddress,superscriptaddress,amsmath,amssymb,prl]{revtex4-2}

\usepackage[utf8]{inputenc}
\usepackage{graphicx}% Include figure files
\usepackage{dcolumn}% Align table columns on decimal
\usepackage{bm}% bold math
\usepackage[%
    colorlinks=true,
    urlcolor=purple,
    linkcolor=magenta,
    citecolor=cyan
]{hyperref}% add hypertext capabilities
\usepackage[mathlines]{lineno}% Enable numbering of text and display math/
\usepackage{siunitx}
\usepackage{amsmath}
\usepackage{physics}
\usepackage{multirow}
\usepackage{kotex}
\usepackage[nodisplayskipstretch]{setspace}

\usepackage[most]{tcolorbox}
\usepackage{xcolor}
\usepackage{soul}

\begin{document}
\title{Demonstrating a Bi-directional Asymmetric Frequency Conversion in Nonlinear Phononic Crystals}
 \author{Yeongtae Jang}\email{These authors contributed equally to this work.}
     \affiliation{%
     Department of Mechanical Engineering, Pohang University of Science and Technology (POSTECH), Pohang 37673, Republic of Korea}
     \email{jrsho@postech.ac.kr}
 \author{Beomseok Oh}\email{These authors contributed equally to this work.}
      \affiliation{%
      Department of Mechanical Engineering, Pohang University of Science and Technology (POSTECH), Pohang 37673, Republic of Korea}
  \author{Eunho Kim}\email{eunhokim@jbnu.ac.kr}
      \affiliation{%
        Division of Mechanical System Engineering, Jeonbuk National University, Jeonju, 54896, Republic of Korea}
        \email{eunhokim@jbnu.ac.kr}
      \affiliation{%
     Department of JBNU-KIST Industry-Academia Convergence Research, Jeonju, 54896, Republic of Korea}
 \author{Junsuk Rho}\email{jrsho@postech.ac.kr}
     \affiliation{%
     Department of Mechanical Engineering, Pohang University of Science and Technology (POSTECH), Pohang 37673, Republic of Korea}
    \email{jrsho@postech.ac.kr}
     \affiliation{%
     Department of Chemical Engineering, Pohang University of Science and Technology (POSTECH), Pohang 37673, Republic of Korea}
    \affiliation{%
    Department of Electrical Engineering, Pohang University of Science and Technology (POSTECH), Pohang 37673, Republic of Korea}
     \affiliation{%
     POSCO-POSTECH-RIST Convergence Research Center for Flat Optics and Metaphotonics, Pohang 37673, Korea}

\begin{abstract}
Beyond the constraints of conservative systems, altering the wave propagation frequency emerges as a crucial factor across diverse physical domains.
This Letter presents a demonstration of bi-directional asymmetric frequency conversion—either upward or downward—depending on the excitation direction in the elastic domain, beyond uni-directional manners.
We numerically and experimentally demonstrate its practical realization in a model system of cylindrical beam crystals, a type of granular crystal characterized by intrinsic local resonance.
This novel wave transport mechanism operates through the interplay among nonlinear contact, spatial asymmetry, and further coupling of the local resonance.
Thanks to the proposed highly tunable architecture, we demonstrated various ways to manipulate wave transport, including tunable frequency conversion.
Given that the local resonance we employ exemplifies the avoided crossings (i.e., strong coupling effect), our work may inspire investigations into diverse physical nonlinear domains that support material/structural resonance.
\end{abstract}
% \pacs{Valid PACS appear here}% PACS, the Physics and Astronomy
\maketitle
\begin{figure}
\includegraphics [width=0.49\textwidth]{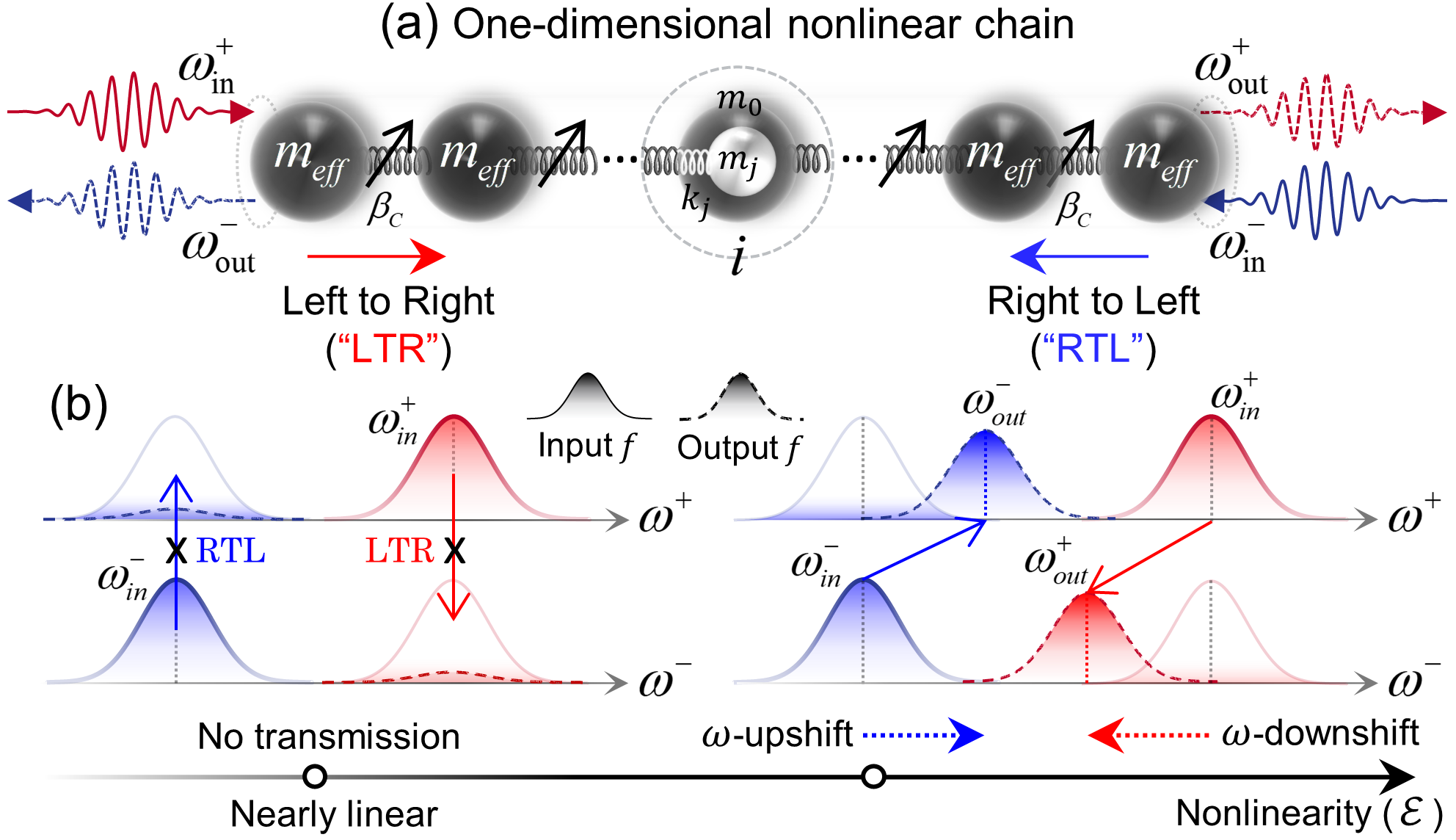}
\caption{\textbf{Bi-directional asymmetric frequency conversion} (a) A one-dimensional locally resonant spring-mass chain, where neighboring cells interact through a nonlinear potential. This nonlinear chain features broken spatial symmetry and local resonance. Gaussian wave-packet propagation in the (+) direction (left to right, LTR) and (-) direction (right to left, RTL). (b) Wave transport scenarios, depending on whether the system exhibits nonlinearity. Solid (dashed) lines indicate the input (output) frequency spectrum. Left: Bi-directional prohibition of wave transmission in the linear regimes. Right: Bi-directional asymmetric wave transfer with frequency conversion in the nonlinear regimes.}
\label{fig1}
\end{figure}

%%%%%%%%%%%%%%%%%%%%%%%%%%%%%%%%%%%%%%%%%%%%%%
One of the fundamental elements of wave propagation is frequency, which dictates how waves interact with matter. 
In a closed linear system, the laws of energy and momentum conservation stemming from time-reversal symmetry prevent the alteration of the assigned frequency.
Therefore, changing the frequency allows us to overcome physical constraints such as reciprocity~\cite{Nassar2020}, the diffraction limit~\cite{Rugar_1984,Zhong2023}, and energy localization~\cite{Flach1998}.
As a result, frequency conversion has garnered considerable attention across diverse wave dynamics domains~\cite{Jain_1996,Tyumenev_2022,Martin_2017}.
\\
\indent
To break the time-invariant property of wave propagation, various principles have been put forward and successfully applied.
Examples range from utilizing external bias~\cite{Wu_2022,Henstridge_2022,Zhang_2024} or time-varying media~\cite{Lee_2018,Apffel_2022}, to exploiting external coupling (such as optomechanical interactions~\cite{Chen_2021}) or nonlinear effects~\cite{nayfeh2008nonlinear,Fejer1994,Stolt2021}.
Among these approaches, leveraging nonlinearity is particularly advantageous because it enables the straightforward tuning of the desired frequency by adjusting the wave amplitude.
\\
\indent
Expanding on basic frequency conversion, recent studies in phononics have demonstrated not only frequency alteration but also simultaneous asymmetric wave transmission using various nonlinear effects~\cite{Lepri2011,Li2011,Zhou2018,Devaux2015,Popa2014,Kim2019}.
These advancements have significantly promoted innovative applications, including acoustic diodes~\cite{Nesterenko2005,Liang2009}, switching~\cite{Boechler2011}, and logic gates~\cite{Li2014}.
Typically, these wave transfers are achieved using structures that introduce internal defects and heterostructures to break spatial symmetry. 
By doing so, the system enables the definitive frequency conversion upward or downward in one direction while inhibiting wave transmission in the opposite direction. 
However, this frequency conversion is limited to a uni-directional scheme.
\\
\indent
Building upon this phenomenon, we posed the question: \textit{Can we realize asymmetric frequency conversion depending on the direction of excitation within the same system?}. 
This pursuit promises more freedom in wave manipulation, driving the motivation behind this study.
Indeed, a similar wave transfer concept was initially proposed for quantum-level nonlinear devices using parametric biharmonic pumps~\cite{Kamal_2014}.
However, achieving this phenomenon requires sophisticated wave mixing, and experimental realization remains a challenge yet to be fully realized.
\\
\indent
In this Letter, we demonstrate bi-directional asymmetric frequency conversion (BiAFC) in phononics by cylindrical beam crystals.
Such systems, known as woodpile structures~\cite{Kim2014,kim2015}, are types of granular crystals~\cite{Chong2017,Nesterenko_2001} made from slender cylindrical particles interacting through the nonlinear Hertz contact law~\cite{johnson1987contact}.
These architectures are tunable and can demonstrate a plethora of wave physics by adjusting the nonlinearity from linear to strongly nonlinear depending on the excitation amplitude~\cite{kim2015,Jang_singular,Kim_2017,Chaunsali2017}.
What makes this system special compared to classical 
granule medium is “local resonance coupling”.
Here, our primary finding is that the interplay among nonlinearity, local resonance, and graded spatial variation enables BiAFC. 
Based on the excitation direction, we show that the assigned frequency underwent up-conversion or down-conversion within the same system.
The unveiled dynamics do not rely on general wave-mixing methods~\cite{Fejer1994,Kauranen2012}, laying the foundation for straightforward energy conversion. 
Remarkably, the converted frequency can be tuned in a controllable manner simply by adjusting the local resonant frequency.
\vspace{1mm}
\\
\indent
\emph{Schematic arguments.}---The underlying mechanism of the BiAFC is schematically depicted in Fig.~\ref{fig1}.
To begin with a general discussion in phononics, we consider a generic locally resonant spring-mass system [Fig.~\ref{fig1}(a)].
%Neighboring cells interact through a nonlinear potential.
The neighboring cells interact through a nonlinear spring indicated by inclined arrows.
This nonlinear chain features broken spatial symmetry along with local resonance.
Gaussian-modulated waveforms are applied as the input, with wave propagation direction set as (+) for left to right (LTR) and (-) for right to left (RTL).
\\
\indent
Wave transport in this system can appear in two distinct ways, reliant on whether the system exhibits nonlinearity: when subjected to small-amplitude excitation, the system is limited to linear dynamics, which results in the bi-directional prohibition of wave transmission [see Fig.~\ref{fig1}(b)].
However, as the amplitude increases, nonlinear interactions between the particles may enable wave transmission.
Notably, the nonlinear effects also give rise to the frequency conversion [see Fig.~\ref{fig1}(c)], which exhibits frequency shifts upward or downward depending on the direction of incidence. 
More specifically, in RTL (LTR) propagation, a Gaussian-modulated pulse centered at $\omega_\text{in}^-$ ($\omega_\text{in}^+$) experiences an up (down) shift, while LTR (RTL) transmission is restricted at $\omega_\text{in}^-$ ($\omega_\text{in}^+$).
This novel wave transport is realized in macroscale phononic crystals, as we demonstrate in the following.
\begin{figure*}[htp]
   \includegraphics [width=0.99\textwidth]{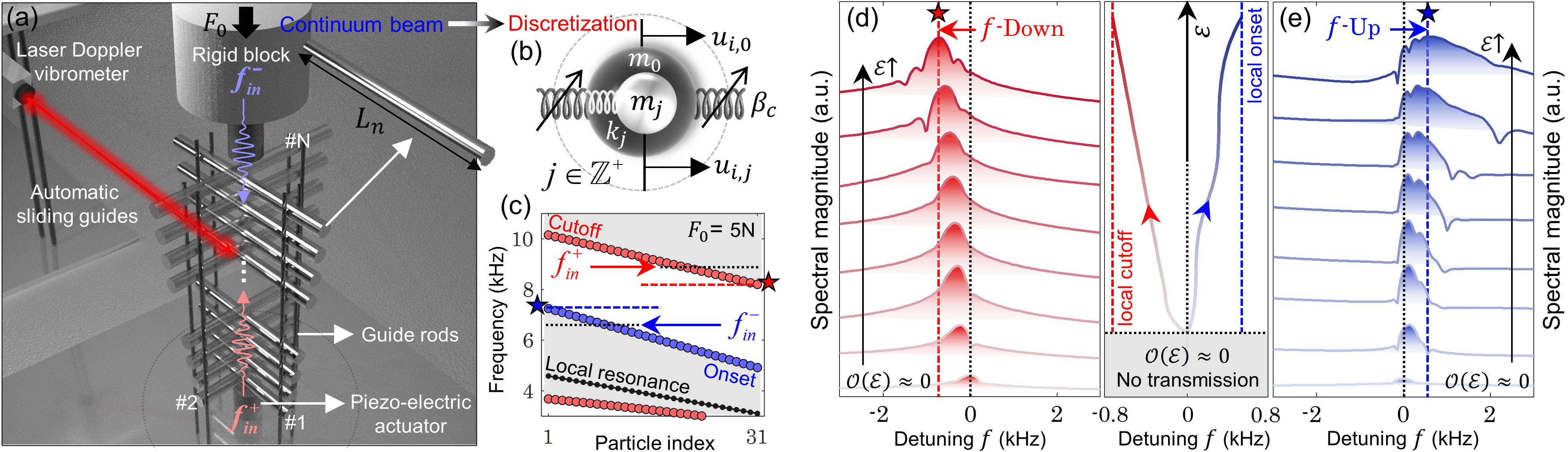}
   \caption{\textbf{Wave dynamics from linear to nonlinear regimes.} (a) Schematic of the experimental setup. (b) Theoretical discretized unit cell representing a single cylindrical element; see the main text for details. (c) Quasi-band diagram obtained from the boundary edges of the dispersion curve for each unit cell. The white area denotes passband regimes, and the gray area denotes the local resonance bandgap. Frequency response as wave amplitude increases for (d) LTR excitation and (e) RTL excitation. As the system's nonlinearity increases, the LTR direction undergoes frequency down-conversion, while the RTL direction exhibits frequency up-conversion (i.e., BiAFC). The converted frequencies converge to the local eigenfrequencies (red and blue stars). The middle panel summarizes the detuning frequency for both upward and downward conversions.}
   \label{fig2}
\end{figure*}
\vspace{1mm}
\\
\indent
\emph{Experimental and Theoretical Setups.}---For the proposed concept, achieving a tunable local resonance within a unit cell is crucial. 
We achieve this by utilizing cylindrical beam crystals as shown in Fig.~\ref{fig2}(a).
In such a system, the local resonance is “intrinsic” (i.e., the vibrational bending mode of the continuum beam itself~\cite{Kim2014}), which enables seamless tuning of the resonant frequencies by adjusting the length ($L$) of the cylindrical beam.
Each cylinder, made of fused quartz, contacts its center orthogonally with the adjacent cylinders.
Our elastic chain consists of 31 cylinders with lengths gradually varying from 90 to 110~mm while maintaining their diameters. 
This spatially asymmetric chain leads to asymmetric local resonance due to the relationship of $\omega\propto{1/L^2}$~\cite{inman1994}.
By applying a pre-compression ($F_0$~=~$5\:\text{N}$), we form a packed ensemble of cylindrical beams.
The guide rods ensure out-of-plane oscillation of the beam.
We then excite a Gaussian-modulated wave into the chain through a piezoelectric actuator for either direction, i.e., LTR ($\omega^{+}_{\text{in}}$) or RTL ($\omega^{-}_{\text{in}}$).
The particle velocity of each beam near the contact point is measured using a laser Doppler vibrometer (Polytec; OFC-534) mounted on a linear stage.
Details of the experiments and data acquisition are provided in the Supplemental Material (SM)~\cite{supp},~Sec.~I.
\\
\indent
The wave dynamics of the cylinder chain along the axes of the contacts are investigated using a discrete element model (DEM). 
In this approach, we represent single cylindrical beams by a primary point mass ($m_0$) with multiple resonators [Fig.~\ref{fig2}(b)].
The internal resonators consist of local spring ($k_j$) and mass ($m_j$), where $j$ represents the number of modes. 
The parameter of a discretized unit cell is obtained using a physics-informed discrete element modeling approach~\cite{Jang_2024}, rooted in continuum beam theory and wave dynamics within periodic structures.
Thus, it simplifies the dynamic calculation process efficiently while maintaining analytical accuracy.
Detailed mathematical modeling is provided in the SM~\cite{supp},~Sec.~II.~A and B.
\\
\indent
The unit cells interact through nonlinear potential, and their behavior is governed by Hertz contact~\cite{johnson1987contact}.
The contact force between the $(i)$-th and $(i+1)$-th beam can be written as $F$=~$\beta_c\left[\delta_i+u_i-u_{i+1}\right]_{+}^{3/2}$, where $\beta_c$ is the stiffness coefficient, and $\delta_i$ is the initial displacement due to the $F_0$; the bracket $[\cdot]_+$=$\text{max}(\cdot,0)$ indicate that it only supports compressive forces, not tensile forces.
As indicated by the contact force formula, the system's nonlinearity is adjusted by varying the dynamic and static displacements: from linear regimes when $|u_{i}-u_{i+1}|\ll\delta_i$ to nonlinear regimes when $|u_{i}-u_{i+1}|\gtrsim\delta_i$ (see the SM~\cite{supp},~Sec.~II.~C for detailed numerical simulations).
\vspace{1mm}
\\
\indent
\emph{Linear spectrum.}---We first begin with the linear dynamics.
Based on the discretized building blocks, it is straightforward to establish a group of local eigenstates (\textit{quasi-band}), calculated from the edges of dispersion curves. 
In the dispersion curve, we initially characterize two distinct local eigenstates that form the \textit{passband}: the onset frequency at $k$=~$0$ (where $k$ is the wavenumber) and the cutoff frequency at $k$=~$\pi$ (see SM~\cite{supp},~Sec.~II).
In Fig.~\ref{fig2}(c), we present a quasi-band diagram of our gradient-index cylinder chain, focusing on the vicinity of the first resonance band.
The shaded gray area represents a local resonance-induced bandgap with locally resonant frequencies indicated by black dots.
We note that this quasi-band originates from the linear limit, accurately describing linear dynamics, but that eigenstates may change in nonlinear regimes.
However, as the system transitions into nonlinear regimes, it can be effectively interpreted in terms of perturbations of this linear basis.
\vspace{1mm}
\\
\indent
\emph{Nonlinear dynamics.}---We now investigate wave dynamics across a range of small (linear) to large (nonlinear) amplitude excitations by numerical simulations.
In our setup, we assign Gaussian-modulated center frequency $f^+_{\text{in}}$ =~$8.92$ kHz and $f^-_{\text{in}}$ =~$6.63$ kHz.
In Figs.~\ref{fig2}(d)-(e), we show the evolution of the frequency spectrum of the output as the system's nonlinearity increases in the LTR and RTL directions.
The plot shown between those in Fig.~\ref{fig2}(d) and \ref{fig2}(e) summarizes the detuning-frequency trends as the system nonlinearity increases.
\\
\indent
In the linear limits, the driven frequency encounters the bandgap in both directions, as depicted in the quasi-band [red and blue arrows in Fig.~\ref{fig2}(c)]. 
This prevents waves from propagating to the end of the chain and inhibits frequency conversion as well.
%Due to our finite systems, a small amplitude of the assigned frequency component can still be detected in the output.
Additionally, it should be noted that spatial gradient systems~\cite{Tsakmakidis2007,Bennetts2018,RomeroGarca2013} generally support \textit{local wave amplification} owing to changes in propagating group velocity from graded eigenmodes (referred to as gradons in various fields~\cite{Gradon11,Gradon22,Gradon33}).
Hence, for the Gaussian-modulated wave input, a boomerang-shaped wave motion is achieved in the spatiotemporal map.
Regarding this wave transfer, comprehensive numerical and experimental findings are provided in the SM~\cite{supp},~Sec.~III (e.g., Figs.~S4, S5, and Movies S1, S2). 
\\
\indent
As the amplitude further increases, the wave dynamics shift to nonlinear regimes.
Therefore, the wave packet can reach the end of the chains beyond linear limits, i.e., the penetration of the bandgap. 
Moreover, a frequency conversion is observable: down-conversion in the LTR direction and up-conversion in the RTL direction, referred to as BiAFC.
Frequency conversion becomes even more evident for large amplitudes.
The full 2D spectrum map is available in the SM~\cite{supp},~Sec.~IV.
Remarkably, the final asymmetric converted frequencies in both directions are saturated at specific local frequencies, indicated by red and blue stars.
These local frequencies are akin to the onset and cutoff frequencies of the first ($\#$1) and last ($\#$31) elements observed in the quasi-band [Fig.~\ref{fig2}(c)].
This is because, despite perturbations owing to nonlinearity, the most dominant spectrum remains on the original basis.
This implies that the input wave packet frequencies can follow the trajectories of the asymmetric local eigenstates in the form of spatially extended modes in nonlinear regimes.
In colloquial terms, the assigned frequencies undergo climbing or descending quasi-bands, which depend on the excitation direction.
\\
\indent
We note that the study of these extended modes (gradons) can be traced back to arguments given in the previous literature~\cite{Kim2019}.
\begin{figure}%[htp]
\includegraphics [width=0.49\textwidth]{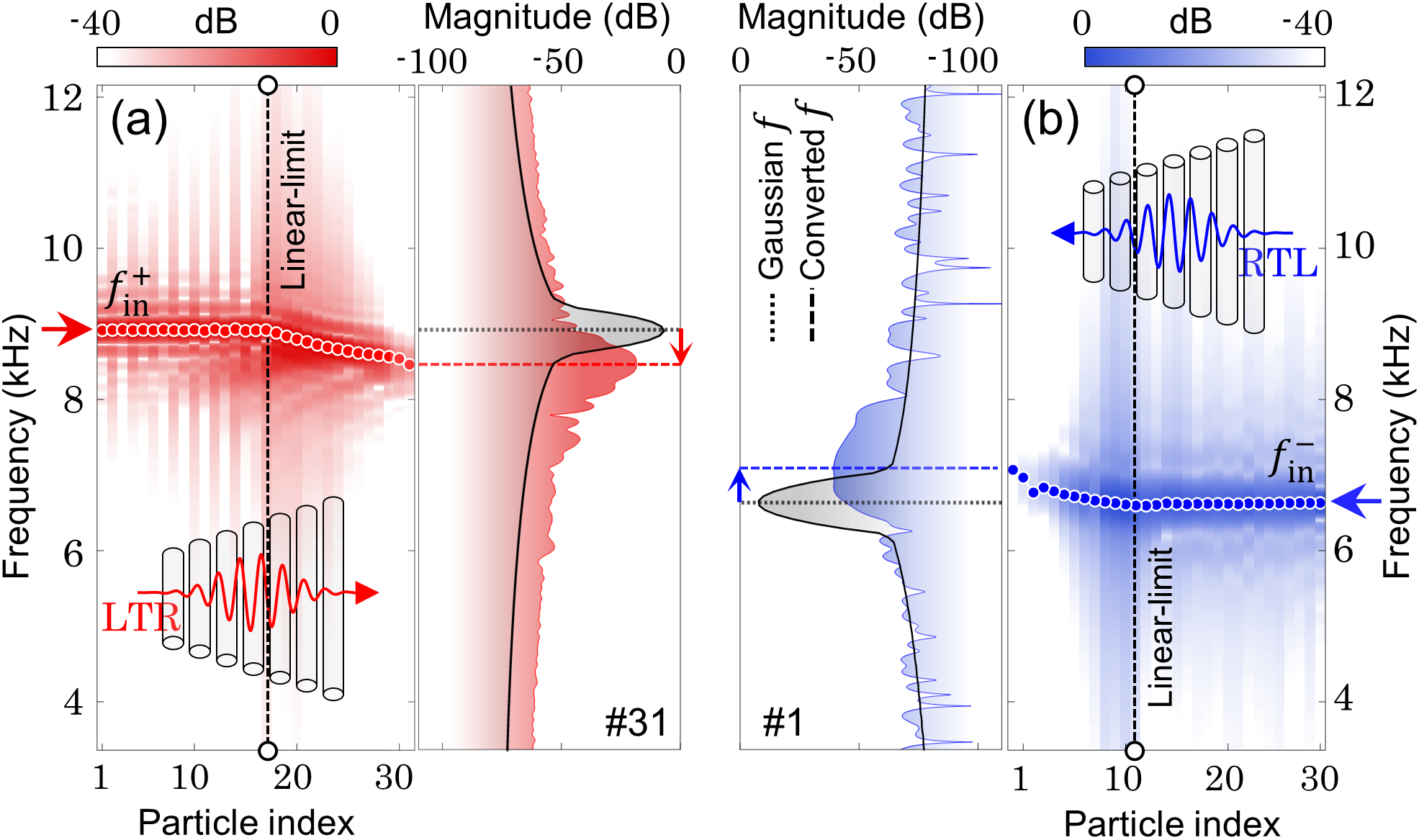}
\caption{\textbf{Experimental validation of bi-directional and asymmetric frequency conversion.} Spatial-frequency map for (a) downward conversion (LTR direction) and (b) upward conversion (RTL direction), obtained by calculating the Fourier transform of particle velocity along the chain. The local peak values of the spectral magnitude are highlighted by circles within the chains. The right panel in (a) and the left panel in (b) display the converted-frequency spectrum of the last cylinders for each direction ($\#$31 for LTR and $\#$1 for RTL).}
\label{fig3}
\end{figure}
This study extensively investigated the asymmetric transmission of waves (i.e., nonreciprocity) in classical gradient-index granular crystals, demonstrating that wave transport undergoes uni-directional frequency down-conversion.
However, the novelty of our work lies in demonstrating bi-directional asymmetric frequency transformation.
This remarkable achievement is attributed to our strategic use of \textit{local resonance} bands.
Thus, in principle, through a more systematic design of the chain configuration, BiAFC could potentially be extended to higher modes due to the semi-infinite modes present in our continuum beam~\cite{inman1994}.
%%%%%%%%%%%%%%%%%%%%%%%%%%%%%%%%%%%%%%%%%%%%%%%%%%%%%%%%%%%%%%%%%%%%%%
\vspace{1mm}
\\
\indent
\emph{Experimental demonstration.}---Now, we experimentally demonstrate this wave phenomenon in our cylindrical beam crystal. 
We excite the first cylinder of the chain in the LTR ($f^+_{\text{in}}$) and RTL ($f^-_{\text{in}}$) directions by driving the actuator with a high amplitude to invoke nonlinearity. 
In Figs.~\ref{fig3}(a)-(b), we show spatial-frequency maps by performing the Fourier transformation of each particle's velocity within the chains.
In these maps, we track the local frequencies with the maximum spectral magnitude (indicated by circles) along the chain.
Even beyond the linear limits established in quasi-bands (dotted line), we observe distinct frequency components. 
In addition, we observe the minimal variation in the input frequencies up to these linear limits. 
This suggests that the system's nonlinearity emerges near the linear limit, which stems from graded spatial features, including large amplitudes. 
Thus, nonlinear effects can cause the extended frequency components to transition to the next eigenstate beyond the bandgap.
\begin{figure}%[!htp]
\includegraphics [width=0.49\textwidth]{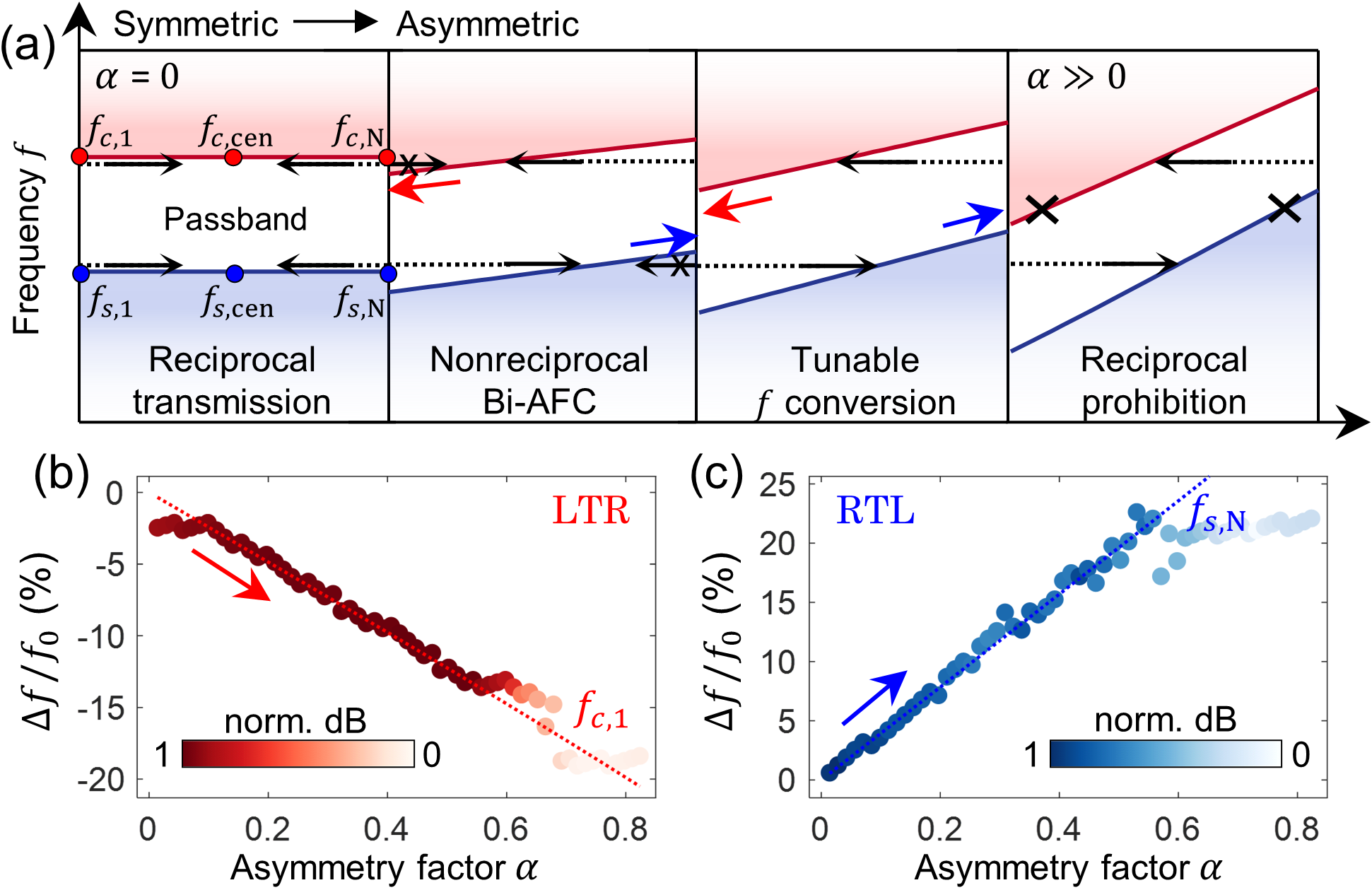}
\caption{\textbf{Effects of the asymmetric chain configuration on frequency conversion.} (a) Schematic illustration of wave transport scenarios to the asymmetry factor $\alpha$. Each panel shows a quasi-band diagram for increasing $\alpha$, with black arrows indicating the excitation frequency and direction. Numerical results of fractional detuning frequency $\Delta{f}/f_0$) and normalized decibel magnitude for (b) down-conversion (LTR) and (c) up-conversion (RTL) as a function of $\alpha$. The red (blue) dotted line indicates the predicted cutoff (onset) frequency $f_{c,1} (f_{s,N})$, at the last cylinder for each configuration.}
\label{fig4}
\end{figure}
This transition process recurs until the wave packet reaches the end of the chain, and this observation supports the previously argued concepts of climbing or descending frequencies.
Therefore, this mechanism exhibits spatially \textit{gradual} frequency “down-conversion” and “up-conversion” depending on the excitation direction.
We further support this argument by performing short-time Fourier transforms on each particle across the entire chain (see the SM.~\cite{supp}, Sec. IV).
\\
\indent
The final converted-frequency spectrum ($f^+_{\text{out}}$ and $f^-_{\text{out}}$) of the last cylinder ($\#$31 and $\#$1) is shown in the right panel of Fig.~\ref{fig3}(a) and the left panel of Fig.~\ref{fig3}(b).
The black lines represent the frequency spectra of excited Gaussian-modulated waves.
We note that the converted bandwidth is relatively preserved because of our system's gradual variation, rather than being defective or heterostructures, which helps minimize wave scattering.
We also observe a difference in conversion bandwidths for BiAFC processes: narrow for downconversion and broader for upconversion.
This difference arises from the nature of the wavenumber, where the onset eigenstate for upconversion yields a broad bandwidth due to the \textit{long}-wavelength, while downconversion associated with \textit{short}-wavelength yields a narrower bandwidth.
In the main text, while our primary focus is on frequency conversion arising from nonlinear effects, we also emphasize our observations in linear dynamics, such as resonance gradons and damping forces featuring wavenumber-dependent behavior, described in SM~\cite{supp}.
\vspace{1mm}
\\
\indent
\emph{Effects of asymmetry.}---Next, we examine key parameters—the system's gradient profile—for tunable frequency conversion in a controllable manner.
Our analytical modeling enables straightforward exploration of various bi-directional control methods.
To examine the influence of the gradient profile, we define an asymmetry factor $\alpha$, which quantifies the slope of the quasi-band profiles, $\alpha$=~$(\alpha_\textnormal{c}+\alpha_\textnormal{s})/2$ with $\alpha_\textnormal{c}$=~$|f_\textnormal{c,1}-f_\textnormal{c,N}|/f_\textnormal{c,cen}$ and $\alpha_\textnormal{s}$=~$|f_\textnormal{s,1}-f_\textnormal{s,N}|/f_\textnormal{s,cen}$; where, $f_\textnormal{c(s),1}$, $f_\textnormal{c(s),cen}$, $f_\textnormal{c(s),N}$ indicate the cutoff (onset) frequencies of first, center, and last cylinder [see Fig.~\ref{fig4}].
To create a linearly sloped eigenstate profile, we fix the eigenstate of the central cylinder.
\\
\indent
In Fig.~\ref{fig4}(a), we depict a schematic of potential wave transmission (and conversion) scenarios based on the degree of slope.
For a symmetric configuration with $\alpha$=~$0$, (i.e., a homogeneous chain), bidirectional wave transmission can occur within the passband under the given excitation [see the black arrow in Fig.~\ref{fig4}(a)] without frequency change. 
Initially, as $\alpha$ increases slightly from 0, BiAFC is achieved following asymmetric eigenstates, and with further elevation of the gradient, tunable frequency conversion becomes possible.
Yet, with $\alpha$ reaching very large, wave transmission, as well as frequency conversion, become unattainable for bidirectional excitation.
This is because of the significant frequency interval between nonlinearity-induced frequencies and neighboring local eigenfrequency.
\\
\indent
To further verify this situation, we perform a full numerical simulation on our given chain by increasing the asymmetry factor.
In Figs.~\ref{fig4}(b)-(c), we present the detuning frequency, expressed as a percentage relative to the excitation frequency, for both LTR and RTL directions.
The color indicates the normalized magnitude of the Fourier intensity. 
As the system's slope increases, we observe a steady increase in the converted frequency in both directions.
Notably, our conversion mechanism shows that the final converted frequency saturates with minimal deviation at $f_{c,1}$ in the LTR direction and at $f_{s,N}$ in the RTL direction. 
When the asymmetry factor becomes very large, the frequency conversion becomes less well-defined for a given excitation amplitude.
%%%%%%%%%%%%%%%%%%%%%%%%%%%%%%%%%%%%%%%%%%%%%%%%%%%%%%%%%%%%%%%%%%%
\vspace{1mm}
\\
\indent
\emph{Conclusion.}---We have demonstrated a novel frequency conversion phenomenon that shows bidirectional asymmetric conversion—either upward or downward—depending on the excitation direction in macro-scale nonlinear cylindrical granules.
Beyond unidirectional frequency conversion schemes, this novel wave transmission arises from the interplay among nonlinear contact, spatial asymmetry, and local resonance coupling.
Our experimental findings show that excited Gaussian-modulated waves undergo spatially gradual frequency downconversion and upconversion without the need for a general wave-mixing method.
Remarkably, the proposed architecture is highly tunable. The contact law allows exploration from linear to nonlinear dynamics, and adjusting the length of cylindrical beams tunes local resonance frequencies. This tunability enables various methods to control elastic energy flow, including frequency conversion.
We believe our findings will open new opportunities for engineering devices with novel bidirectional functionalities, relevant to applications such as vibration filtering, rectification, logic gates, and signal processing.
Furthermore, given that the local resonance coupling we utilize is a well-established example of avoided crossings~\cite{Novotny2010}, we expect our analysis to extend to various physical domains with material/structural resonances, including polaritons~\cite{Basov2016}, photonics, and metamaterials.
%%%%%%%%%%%%%%%%%%%%%%%%%%%%%%%%%%%%%%%%%%%%%
\section*{Acknowledgements}
This work was financially supported by the POSCO-POSTECH-RIST Convergence Research Center program funded by POSCO, and the National Research Foundation (NRF) grants (RS-2024-00356928, NRF-2019R1A5A8080290) funded by the Ministry of Science and ICT (MSIT) of the Korean government, and the grant (PES4400) from the endowment project of “Development of smart sensor technology for underwater environment monitoring” funded by Korea Research Institute of Ships Ocean engineering (KRISO). "E.K. also acknowledges the support of the National Research Foundation grant (NRF-2020R1A2C2013414), the Commercialization Promotion Agency for R\&D Outcomes (COMPA) grant funded by the Korean Government (Ministry of Science and ICT, 2023), and the KIST Institutional Program (Project No. 2Z07050-24-P035). B.O. acknowledges the NRF Ph.D. fellowship (RS-2024-00409956) funded by the Ministry of Education of the Korean government.
\bibliographystyle{apsrev4-2}
\bibliography{Reference.bib}

\end{document}